\documentclass{PoS}
\usepackage{psfrag,epsfig,graphicx,graphics}
\usepackage{wrapfig}
\usepackage{amsmath}

\def\slashchar#1{\setbox0=\hbox{$#1$}
   \dimen0=\wd0
   \setbox1=\hbox{/} \dimen1=\wd1
   \ifdim\dimen0>\dimen1
      \rlap{\hbox to \dimen0{\hfil/\hfil}}
      #1
   \else
      \rlap{\hbox to \dimen1{\hfil$#1$\hfil}}
      /
   \fi}

\def\bei{\begin{itemize}}
\def\ei{\end{itemize}}

\def\beeq{\begin{eqnarray}} 
\def\beqa{\begin{eqnarray}}
\def\bea{\begin{eqnarray}}

\def\eea{\end{eqnarray}}
\def\eqa{\end{eqnarray}}
\def\eeeq{\end{eqnarray}}

\def\eqar{\end{array}}
\def\beqar{\begin{array}}

\def\beas{\begin{eqnarray*}}
\def\beqas{\begin{eqnarray*}}

\def\eqas{\end{eqnarray*}}
\def\eeas{\end{eqnarray*}}

\def\beq{\begin{equation}} 
\def\be{\begin{equation}}

\def\ee{\end{equation}}
\def\eq{\end{equation}}
\def\eeq{\end{equation}}

\def\beqd{\begin{displaymath}}
\def\eeqd{\end{displaymath}}
\def\eqd{\end{displaymath}}

\def\beeq{\begin{eqnarray}} \def\eeeq{\end{eqnarray}}

\newcommand{\fin}{\end{document}}

\newcommand{\veck}{{\bf k}}

\newcommand{\veckone}{{\bf k}_1}
\newcommand{\vecktwo}{{\bf k}_2}
\newcommand{\veckj}{{\bf k}_{J}}

\newcommand{\veckjone}{{\bf k}_{J,1}}
\newcommand{\veckjtwo}{{\bf k}_{J,2}}

\newcommand{\deins}[1]{{\rm d}#1\,}
\newcommand{\dzwei}[1]{{\rm d}^2#1\,}
\newcommand{\dk}{\dzwei{\veck}}

\newcommand{\dkone}{\dzwei{\veckone}}
\newcommand{\dktwo}{\dzwei{\vecktwo}}

\newcommand{\dsigma}{\deins{\sigma}}
\newcommand{\dsigmahat}{\deins{{\hat\sigma}_{\rm{ab}}}}
\newcommand{\dnu}{\deins{\nu}}

\newcommand{\dx}{\deins{x}}
\newcommand{\dxone}{\deins{x_1}}
\newcommand{\dxtwo}{\deins{x_2}}
\newcommand{\dyjetone}{\deins{y_{J,1}}}
\newcommand{\dyjettwo}{\deins{y_{J,2}}}
\newcommand{\dphij}{\deins{\phi_{J}}}
\newcommand{\dphijone}{\deins{\phi_{J,1}}}
\newcommand{\dphijtwo}{\deins{\phi_{J,2}}}
\newcommand{\dtwojets}{{\rm d}|\veckjone|\,{\rm d}|\veckjtwo|\,\dyjetone \dyjettwo}

\newcommand{\shat}{{\hat s}}
\newcommand{\non}{\nonumber\\}
\newcommand{\asbar}{{\bar{\alpha}}_s}

\newcommand{\init}{\text{init}}
\newcommand{\MOM}{\text{MOM}}
\newcommand{\MSbar}{\overline{\text{MS}}}
\newcommand{\BLM}{\text{BLM}}

\newcommand{\avgcosn}{\langle \cos n \varphi \rangle}
\newcommand{\avgcosm}{\langle \cos m \varphi \rangle}
\newcommand{\avgcos}{\langle \cos \varphi \rangle}
\newcommand{\avgcostwo}{\langle \cos 2 \varphi \rangle}

\title{Confronting BFKL dynamics with experimental studies of Mueller-Navelet jets at the LHC}

\ShortTitle{Confronting BFKL dynamics with experimental studies of Mueller-Navelet jets at the LHC}

\author{\speaker{Bertrand Duclou\'e}\\
        LPT, Universit\'e Paris-Sud, CNRS, 91405, Orsay, France\\
        E-mail: \email{Bertrand.Ducloue@th.u-psud.fr}}

\author{Lech Szymanowski\\
        National Centre for Nuclear Research (NCBJ), Warsaw, Poland\\
        E-mail: \email{Lech.Szymanowski@fuw.edu.pl}}

\author{Samuel Wallon\\
        LPT, Universit{\'e} Paris-Sud, CNRS, 91405, Orsay, France\\
        UPMC Univ. Paris 06, Facult\'e de Physique, 4 place Jussieu, 75252 Paris Cedex 05, France
        E-mail: \email{Samuel.Wallon@th.u-psud.fr}}

\abstract{The study of the production of Mueller-Navelet jets at hadron colliders, characterized as two forward jets separated by a large interval of rapidity, is known to be one of the best possible tests of the high energy dynamics of QCD. We analyze this process within a complete next-to-leading logarithm framework \`a la BFKL. In addition, we use the Brodsky-Lepage-Mackenzie procedure, here extended to the perturbative Regge dynamics, to fix the renormalization scale to its optimal value. The obtained results provide a very good description of the recent CMS data at the LHC for the azimuthal correlations of the jets.}

\FullConference{XXII. International Workshop on Deep-Inelastic Scattering and Related Subjects \\
		 28 April - 2 May 2014 \\
		 Warsaw, Poland}

\begin{document}

\psfrag{Y}[][][1]{$Y$}

\psfrag{BLM}[l][l][.7]{NLL, $\mu_R=\mu_{R,\BLM}$}
\psfrag{NLL}[l][l][.7]{NLL, $\mu_R=\mu_{R,\init}$}
\psfrag{CMS}[l][l][.7]{CMS data}
\psfrag{Dijet}[l][l][.7]{NLO fixed-order}

\psfrag{dist}[][][1]{$\frac{1}{\sigma}\frac{d \sigma}{d\varphi}$}
\psfrag{phi}[][][1]{$\varphi$}

\section{Introduction}

The high energy dynamics of
QCD, described by the Balitsky-Fadin-Kuraev-Lipatov (BFKL)
approach~\cite{Fadin:1975cb,Kuraev:1976ge,Kuraev:1977fs,Balitsky:1978ic}, have been much studied since four decades. One of the most promising processes is the 
production of two forward jets separated by a large interval of rapidity at 
hadron colliders, as proposed by Mueller and Navelet~\cite{Mueller:1986ey}. We here report on our study of this process in a next-to-leading logarithmic (NLL) BFKL approach, which we confront with 
the most recent LHC data extracted by the CMS collaboration
for the azimuthal correlations of these jets~\cite{CMS-PAS-FSQ-12-002}, obtaining a very satisfactory description
 within this framework.

Two main building blocks are involved in the BFKL treatment:
 the jet 
vertex, which describes the transition from an incoming parton to a jet,
and the Green's function, which describes the pomeron exchange between the vertices. A complete NLL BFKL analysis of Mueller-Navelet jets, including the NLL corrections both to the Green's function~\cite{Fadin:1998py,Ciafaloni:1998gs} and to
the jet vertex~\cite{Bartels:2001ge,Bartels:2002yj}, showed that the NLL corrections to the jet vertex have a very large effect, leading to a lower cross section and a much larger azimuthal correlation~\cite{Colferai:2010wu}. Furthermore, this study showed that the results were very dependent on the choice
of the scales, especially the renormalization scale $\mu_R$ and the
factorization scale $\mu_F$, a fact which remains true when using realistic kinematical cuts for LHC experiments~\cite{Ducloue:2013hia}. A way to include higher order contributions in order to reduce this dependency, in a 
 physically motivated way, was proposed by Brodsky, Lepage and Mackenzie (BLM)~\cite{Brodsky:1982gc}. We adhere to this procedure which allows to fix the renormalization scale, 
adapted here
to the resummed perturbation theory \`a la BFKL~\cite{Brodsky:1998kn,Brodsky:2002ka}. Details can be found in ref.~\cite{Ducloue:2013bva}. We also discuss the relevance of energy-momentum conservation in our NLL BFKL treatment.

\section{Basic formulas}

The differential cross-section for the production of two jets of transverse momenta $\veckjone$, $\veckjtwo$ and rapidities
$y_{J,1}$, $y_{J,2}$ is
\begin{equation}
  \frac{\dsigma}{\dtwojets} = \sum_{{\rm a},{\rm b}} \int_0^1 \dxone \int_0^1 \dxtwo f_{\rm a}(x_1) f_{\rm b}(x_2) \frac{\dsigmahat}{\dtwojets},
\end{equation}
where $f_{\rm a, b}$ are the usual collinear partonic distributions. In the BFKL framework, the partonic cross-section reads
\begin{equation}
  \frac{\dsigmahat}{\dtwojets} = \int \dphijone\dphijtwo\int\dkone\dktwo V_{\rm a}(-\veckone,x_1)\,G(\veckone,\vecktwo,\shat)\,V_{\rm b}(\vecktwo,x_2),\label{eq:bfklpartonic}
\end{equation}
where $V_{\rm a, b}$ and $G$ are respectively the jet vertices and the Green's function. Besides the cross-section, the azimuthal correlation is of interest~\cite{DelDuca:1993mn,Stirling:1994zs}. Denoting $\phi_{J,1}$, $\phi_{J,2}$ the azimuthal angles of the two jets, and defining
the relative azimuthal angle $\varphi$ such that $\varphi=0$ corresponds
to the back-to-back configuration, the moments of this distribution read
\begin{equation}
  \langle\cos(n\varphi)\rangle \equiv \langle\cos\big(n(\phi_{J,1}-\phi_{J,2}-\pi)\big)\rangle = \frac{\mathcal{C}_n}{\mathcal{C}_0} \,,
\end{equation}
with
\begin{equation}   
\mathcal{C}_0 = \frac{\dsigma}{\dtwojets} \,,
\end{equation}
and
\begin{equation}
  \mathcal{C}_n = (4-3\delta_{n,0}) \int \dnu C_{n,\nu}(|\veckjone|,x_{J,1})C^*_{n,\nu}(|\veckjtwo|,x_{J,2}) \left( \frac{\shat}{s_0} \right)^{\omega(n,\nu)}\,.
  \label{Cn}
\end{equation}
The coefficients $C_{n,\nu}$ are given by
\begin{equation}
   C_{n,\nu}(|\veckj|,x_{J})= \int\dphij\dk \dx f(x) V(\veck,x) E_{n,\nu}(\veck) \cos(n\phi_J)\,,
  \label{Cnnu}
\end{equation}
where 
\begin{equation}
  E_{n,\nu}(\veck) = \frac{1}{\pi\sqrt{2}}\left(\veck^2\right)^{i\nu-\frac{1}{2}}e^{in\phi}\,.
\label{def:eigenfunction}
\end{equation}
At leading logarithmic (LL) accuracy, the jet vertex reads
\begin{equation}
  V_{\rm a}(\veck,x)=V_{\rm a}^{(0)}(\veck,x) = \frac{\alpha_s}{\sqrt{2}}\frac{C_{A/F}}{\veck^2} \delta\left(1-\frac{x_J}{x}\right)|\veckj|\delta^{(2)}(\veck-\veckj)\,,
\end{equation}
where $C_A=N_c=3$ corresponds to the case of incoming gluon and $C_F=(N_c^2-1)/(2N_c)=4/3$ corresponds to the case of incoming quark. The expressions of the next-to-leading order (NLO) corrections to $V_{\rm a}$~\cite{Bartels:2001ge,Bartels:2002yj}, which have been recently reobtained using various methods in refs.~\cite{Caporale:2011cc,Hentschinski:2011tz,Chachamis:2012cc}, can be found in ref.~\cite{Colferai:2010wu}. They have been computed in the limit of small cone jets in ref.~\cite{Ivanov:2012ms} and used in refs.~\cite{Caporale:2012ih,Caporale:2013uva}. The LL BFKL trajectory reads 
\begin{equation}
  \omega(n,\nu) = \asbar \chi_0\left(|n|,\frac{1}{2}+i\nu\right), \quad \chi_0(n,\gamma) = 2\Psi(1)-\Psi\left(\gamma+\frac{n}{2}\right)-\Psi\left(1-\gamma+\frac{n}{2}\right)\,,
\end{equation}
where $\asbar = N_c\alpha_s/\pi$, while at NLL, it is modified as~\cite{Kotikov:2000pm,Kotikov:2002ab,Ivanov:2005gn,Vera:2006un,Vera:2007kn,Schwennsen:2007hs}
\begin{equation}
  \omega(n,\nu) = \asbar \chi_0\left(|n|,\frac{1}{2}+i\nu\right) + \asbar^2 \left[ \chi_1\left(|n|,\frac{1}{2}+i\nu\right)-\frac{\pi b_0}{N_c}\chi_0\left(|n|,\frac{1}{2}+i\nu\right) \ln\frac{|\veckjone|\cdot|\veckjtwo|}{\mu_R^2} \right]\,,
\end{equation}
where $b_0=\beta_0/(4 \pi)$ with $\beta_0=(11N_c-2N_f)/3$, $N_f$ being the number of flavors. The expression for $\chi_1$, which was derived in
refs.~\cite{Kotikov:2000pm,Kotikov:2002ab}, can be found in eq.~(2.17) of ref.~\cite{Ducloue:2013hia}.
As stated in the Introduction, it was observed that, even at NLL accuracy, several observables depend strongly on the choice of the scales, and in particular the renormalization scale $\mu_R$.
A way to reduce this dependency is to use an optimization procedure to fix the renormalization scale. The BLM procedure~\cite{Brodsky:1982gc}, which we use, is a way of absorbing the non conformal
terms of the perturbative series in a redefinition of the coupling constant, to
improve the convergence of the perturbative series.
Note that the BLM procedure was later extended to all orders, leading to the principle of maximal conformality (PMC)~\cite{Brodsky:2011ig,Brodsky:2011ta,Brodsky:2012rj,Brodsky:2012ik,Mojaza:2012mf,Wu:2013ei,Brodsky:2013vpa,Zheng:2013uja}.
The first practical implementation of the BLM procedure in the context of BFKL was performed in refs.~\cite{Brodsky:1998kn,Brodsky:2002ka}. Here the authors argued that, when dealing with BFKL calculations, the BLM procedure is more conveniently applied in a physical renormalization scheme like the $\MOM$ scheme instead of the usual $\MSbar$ scheme.
This method was followed in
refs.~\cite{Angioni:2011wj,Hentschinski:2012kr,Hentschinski:2013id}.
The observables we have introduced previously in the $\MSbar$ scheme can be obtained in the  $\MOM$ scheme using~\cite{Celmaster:1979dm,Celmaster:1979km}
\begin{equation}
  \alpha_{\MSbar}=\alpha_{\MOM}\left(1+\alpha_{\MOM}\frac{T_{\MOM}}{\pi}\right)\,,
\end{equation}
where $T_{\MOM}=T_{\MOM}^\beta+T_{\MOM}^{conf}$,
\begin{eqnarray}
   T_{\MOM}^{conf} &=& \frac{N_c}{8}\left[\frac{17}{2}I+\frac{3}{2}\left(I-1\right)\xi+
   \left(1-\frac{1}{3}I\right)\xi^2-\frac{1}{6}\xi^3\right], \non
   T_{\MOM}^\beta &=& -\frac{\beta_0}{2} \left(1+\frac{2}{3}I\right),
\end{eqnarray}
where  $I=-2\int_0^1 dx \ln(x)/[x^2-x+1] \simeq 2.3439$ and $\xi$ is a gauge parameter.
After performing the transition to the $\MOM$ scheme, one should then choose the renormalization scale to make the $\beta_0$-dependent part vanish. In the present case, this is achieved with
\begin{equation}
  \mu^2_{R,\BLM}=|\veckjone|\cdot|\veckjtwo| \exp \left[ \frac{1}{2}
  \chi_0(n,\gamma)-\frac{5}{3}+2\left(\!1+\frac{2}{3}I\!\right)\! \right]\!.
\end{equation}

\section{Results: symmetric configuration}

In this section we compare our results with the measurement performed by the CMS collaboration on the azimuthal correlations of Mueller-Navelet jets at the LHC at a center of mass energy $\sqrt{s}=7$ TeV~\cite{CMS-PAS-FSQ-12-002}.
For this we consider two jets with transverse momenta larger than $35$ GeV and rapidities lower than $4.7$. We use the anti-$k_t$ jet algorithm~\cite{Cacciari:2008gp} with a size parameter $R=0.5$ and the MSTW 2008~\cite{Martin:2009iq} parametrization for the parton distribution functions.
On the plots we show the CMS data (black dots with error bars), the NLL BFKL result using the ``natural'' scale choice $\mu_R=\sqrt{|\veckjone| \cdot |\veckjtwo|}$ (solid black line) and the NLL BFKL results using the BLM scale setting (gray error band). The gray error band for this second treatment corresponds to  the typical theoretical uncertainty when practically implementing the BLM procedure.

\begin{figure}[h]
  \begin{minipage}{0.49\textwidth}
    \psfrag{cos}[l][l][.8]{$\avgcos$}
    \includegraphics[width=7.5cm]{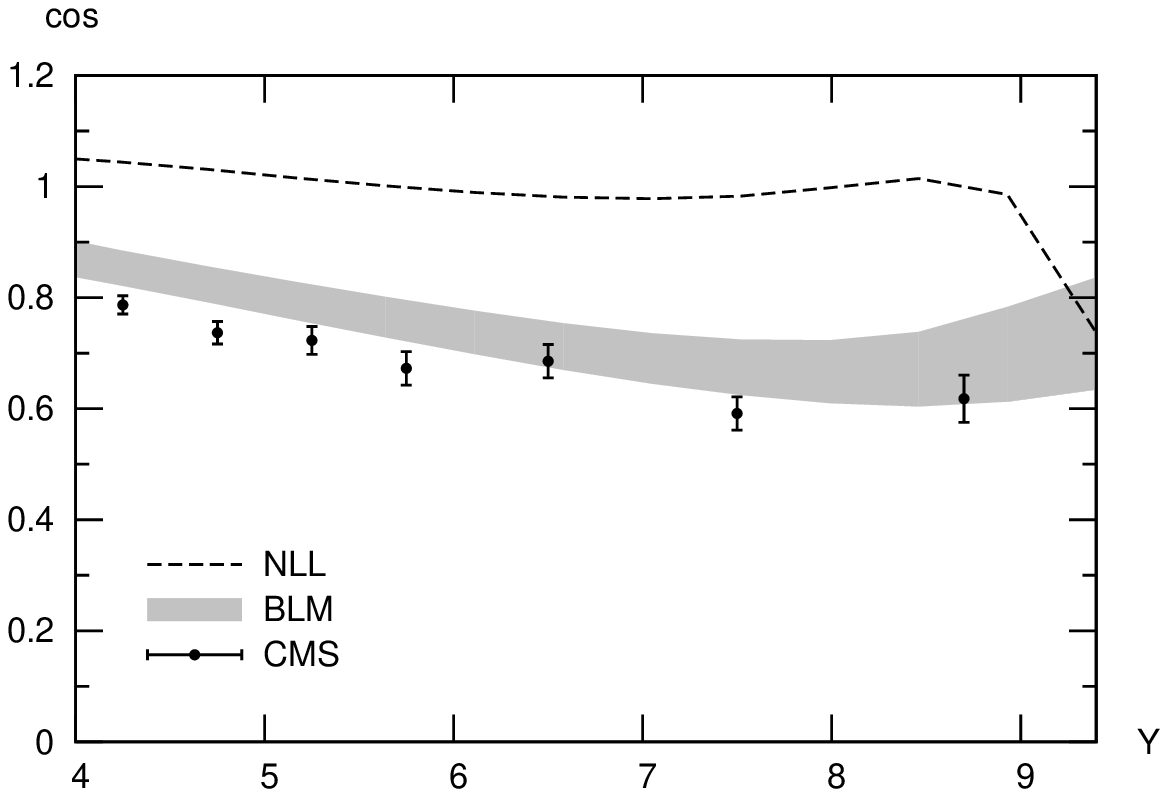}
  \end{minipage}
  \begin{minipage}{0.49\textwidth}
    \psfrag{cos}[l][l][.8]{$\avgcostwo$}
    \includegraphics[width=7.5cm]{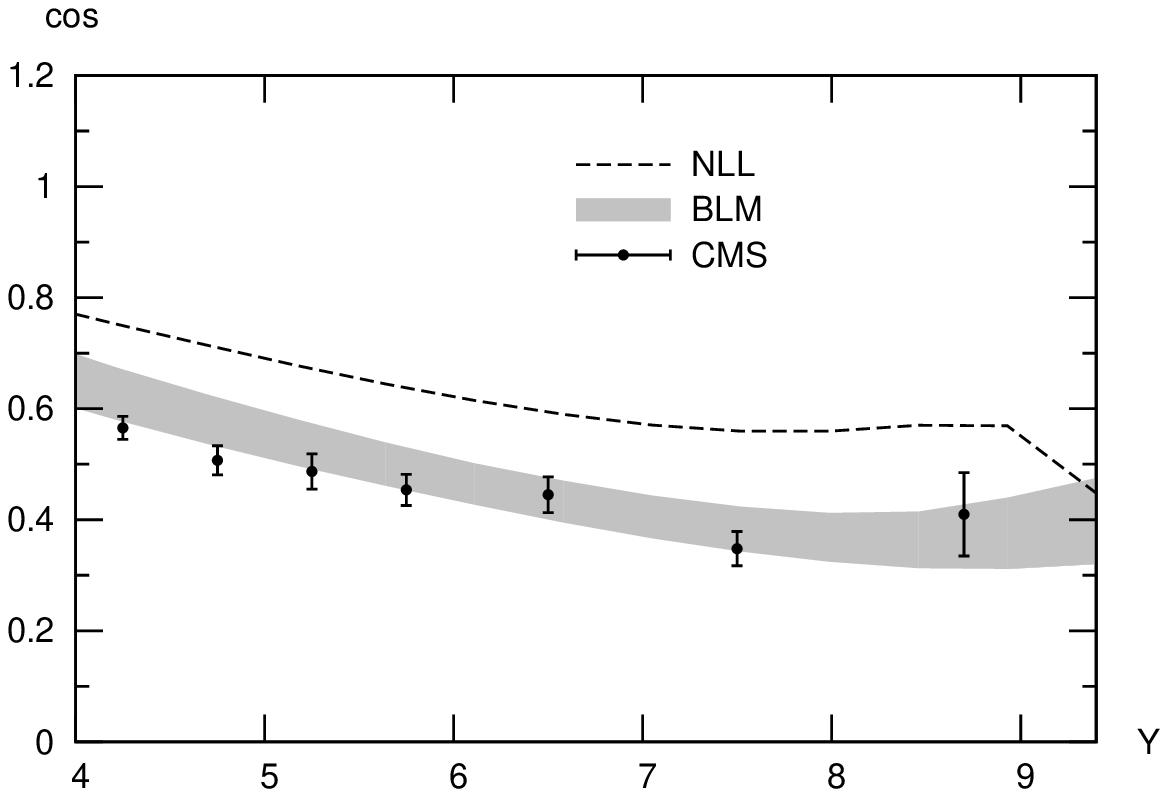}
  \end{minipage}
  \caption{Left: Variation of $\avgcos$ as a function of $Y$ at NLL accuracy compared with CMS data. Right:
  Variation of $\avgcostwo$ as a function of $Y$ at NLL accuracy compared with CMS data.}
\label{Fig:cos-cos2_blm_sym}
\end{figure}

We first show results for the angular correlations $\avgcos$ and $\avgcostwo$ as a function of $Y$ on fig.~\ref{Fig:cos-cos2_blm_sym} (L) and (R) respectively.
The conclusion for these two observables is the same: using the ``natural'' scale choice, the NLL BFKL calculation predicts a too strong correlation. After using the BLM procedure to fix the renormalization scale, the agreement with the data becomes much better.

\begin{figure}
  \begin{minipage}{0.49\textwidth}
    \includegraphics[height=6cm]{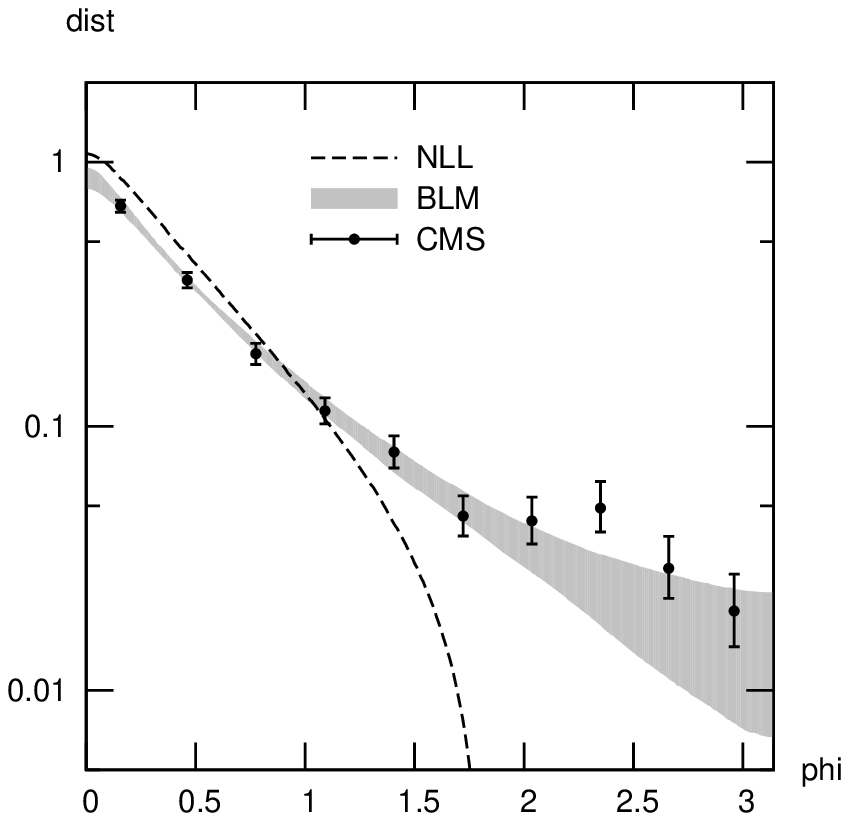}
  \end{minipage}
  \begin{minipage}{0.49\textwidth}
    \psfrag{cos}[l][l][.8]{$\avgcostwo/\avgcos$}
    \includegraphics[width=7.5cm]{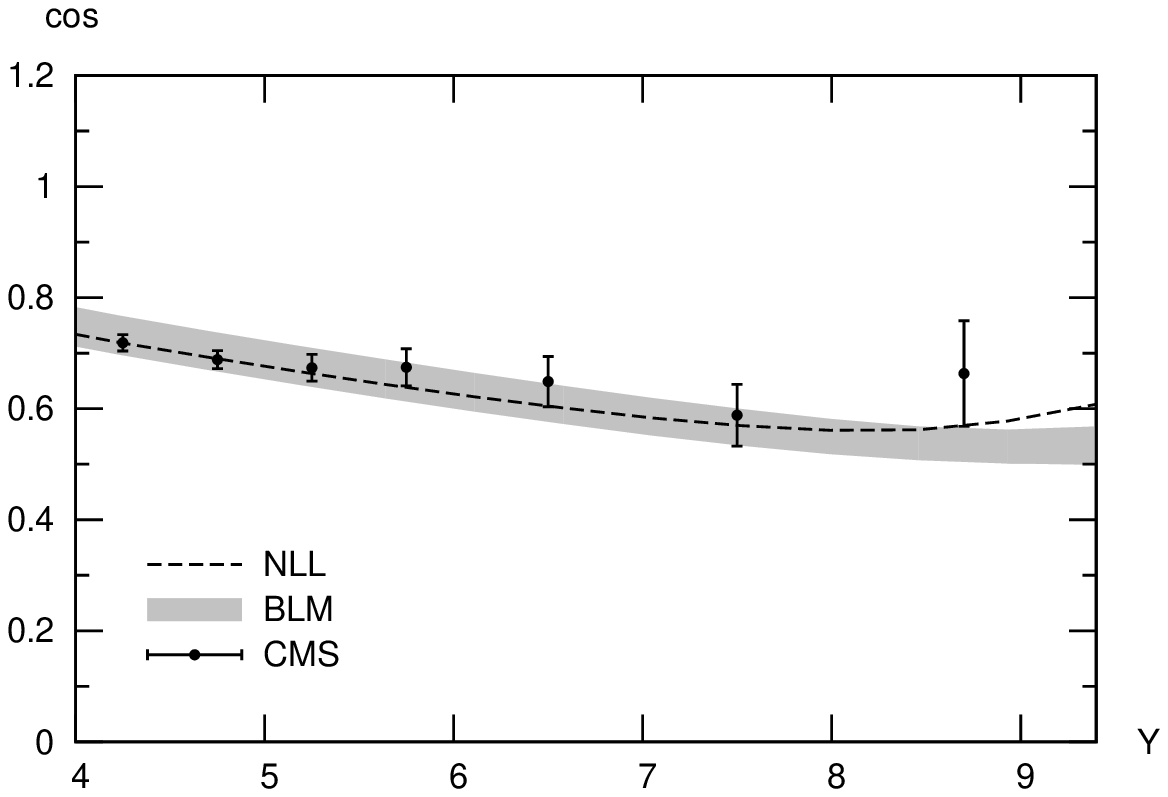}
  \end{minipage}
  \caption{Left: Azimuthal distribution at NLL accuracy compared with CMS data.
  Right: Variation of $\avgcostwo / \avgcos$ as a function of $Y$ at NLL accuracy compared with CMS data.}
\label{Fig:cos2cos-dist_blm_sym}
\end{figure}

This improvement due to the BLM procedure can also be seen from fig.~\ref{Fig:cos2cos-dist_blm_sym} (L), where we show the azimuthal distribution of the jets $\frac{1}{{\sigma}}\frac{d{\sigma}}{d \varphi}$,
which can be expressed as
\begin{equation}
 \frac{1}{{\sigma}}\frac{d{\sigma}}{d \varphi}
  ~=~ \frac{1}{2\pi}
  \left\{1+2 \sum_{n=1}^\infty \cos{\left(n \varphi\right)}
  \left<\cos{\left( n \varphi \right)}\right>\right\}.
\end{equation}

On the other hand, as was already observed both at LL and NLL accuracy~\cite{Vera:2006un,Vera:2007kn,Schwennsen:2007hs,Colferai:2010wu,Ducloue:2013hia}, ratios of the kind $\avgcosm/\avgcosn$ with $n \neq 0$ are much more stable with respect to the scales than individual moments $\avgcosn$ and therefore are almost not affected by the BLM procedure.
This is shown on fig.~\ref{Fig:cos2cos-dist_blm_sym} (R) for $\avgcostwo/\avgcos$, where we see that the good agreement with the data obtained when using the ``natural'' scale choice is still present after applying the BLM procedure.

\section{Results: asymmetric configuration}

The configuration chosen by the CMS collaboration in ref.~\cite{CMS-PAS-FSQ-12-002} does not allow to perform a comparison with a fixed order calculation since these calculation are unstable when the lower cut on the transverse momenta of both jets is the same~\cite{Andersen:2001kta,Fontannaz:2001nq}. Nevertheless, comparing the agreement of a fixed order calculation and of a BFKL one with data would be very useful to study the need to take into account resummation effects at high energy.
In this section, we will compare these two approaches in a slightly different configuration, where the lower cut on the transverse momenta of the jets is not the same. In practice, we use the same cuts as in the previous section but we add the requirement that the transverse momentum of at least one jets is larger than $50$ GeV. A fixed order calculation should give trustable results with these cuts, which could be easily implemented by experimental collaborations.
\begin{figure}[h]
  \psfrag{BLM}[l][l][.7]{NLL BFKL, $\mu_R=\mu_{R,\BLM}$}
  \psfrag{NLL}[l][l][.7]{NLL BFKL, $\mu_R=\mu_{R,\init}$}
  \psfrag{cos}[l][l][.8]{$\avgcostwo/\avgcos$}
  \centering\includegraphics[width=7.5cm]{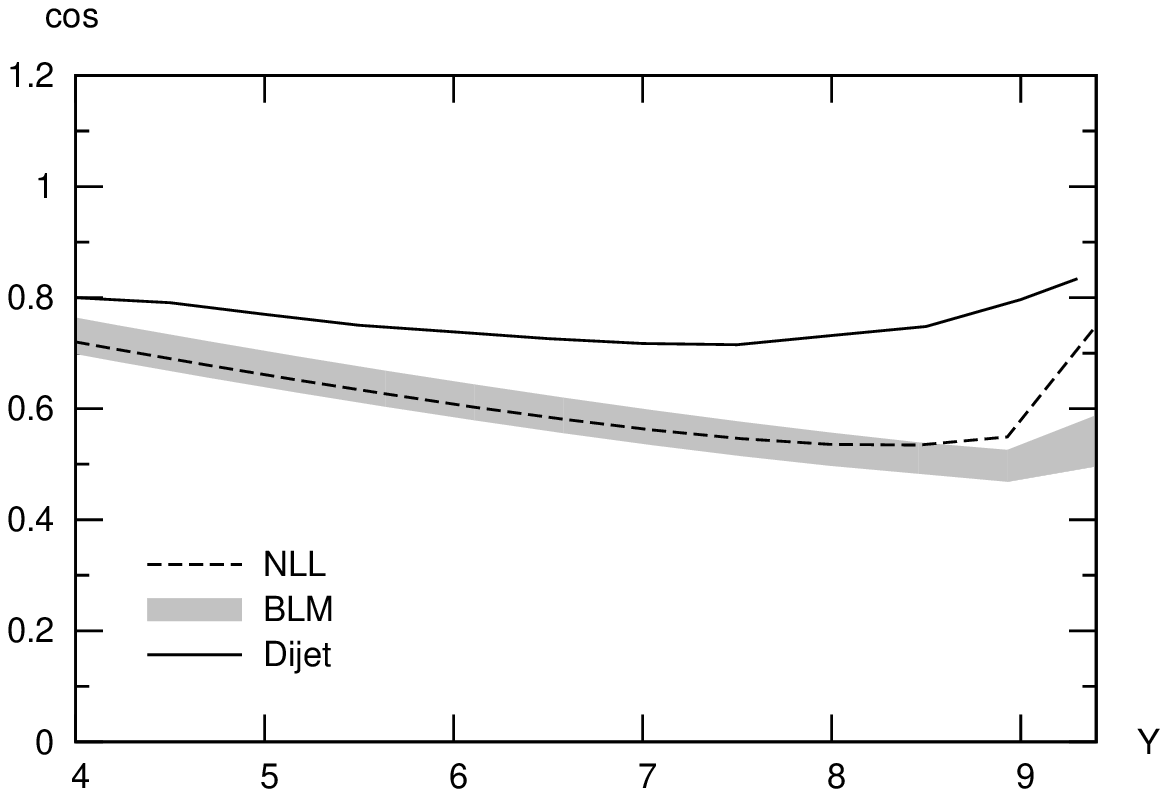}
  \caption{Variation of $\avgcostwo / \avgcos$ as a function of $Y$ at NLL
  accuracy compared with a fixed order treatment.}
\label{Fig:cos2cos_blm_asym}
\end{figure}
As we have discussed in the previous section, the quantities $\avgcosn$ are not very stable even at NLL accuracy in the BFKL approach, therefore the comparison with a fixed order calculation for these observables would not be very meaningful.
On the contrary, we have seen that the observable $\avgcostwo/\avgcos$ is more stable in the BFKL approach. On fig.~\ref{Fig:cos2cos_blm_asym}, we show the comparison of the NLL BFKL calculation with the results obtained with the NLO fixed order code {\textsc{Dijet}}~\cite{Aurenche:2008dn}.
We see that there is a sizable difference between the two treatments over a large $Y$ range.

\section{Energy-momentum conservation}

Even if the above discussed asymmetric configuration is needed to obtain trustable results in the fixed order approach, this could be problematic for the BFKL calculation because of the issue of energy-momentum conservation.
This is a formally sub-leading effect in the BFKL approach, but it was suggested that these effects could be numerically important, at least at LL accuracy.
In particular, the authors of ref.~\cite{DelDuca:1994ng} proposed to evaluate the importance of this effect by comparing the results of an exact $\mathcal{O}(\alpha_s^3)$ calculation with the BFKL result, expanded in powers of $\alpha_s$ and truncated to order $\alpha_s^3$. They found that a LL BFKL calculation strongly overestimates the cross section with respect to an exact calculation as long as the two jets transverse momenta are not very similar (which is the case in the asymmetric configuration discussed in the previous section).
In the same spirit, a study with LO vertices and NLL Green's function was performed in ref.~\cite{Marquet:2007xx}.
Here we will follow the same approach, but we will also include NLO corrections to the jet vertices to see if they lead to a less severe violation of energy-momentum conservation~\cite{us}.
In details, the authors of ref.~\cite{DelDuca:1994ng} introduced an effective rapidity $Y_{\rm eff}$ as
\begin{equation}
  Y_{\rm eff} \equiv\ Y \frac{\mathcal{C}_m^{2\to3}}{\mathcal{C}_m^{{\rm BFKL},\mathcal{O}(\alpha_s^3)}} \,.
  \label{eq:Cm_e-m_cons}
\end{equation}
where $\mathcal{C}_m^{2\to3}$ is the exact $\mathcal{O}(\alpha_s^3)$ results obtained by studying the reaction $gg \to ggg$, while\linebreak ${\mathcal{C}_m^{{\rm BFKL},\mathcal{O}(\alpha_s^3)}}$ is the BFKL result expanded in powers of $\alpha_s$ and truncated to order $\mathcal{O}(\alpha_s^3)$.
The definition of the effective rapidity~(\ref{eq:Cm_e-m_cons}) is motivated by the observation that if one replaces $Y$ by $Y_{\rm eff}$ in the BFKL calculation, expands in powers of $\alpha_s$ and truncates to order $\alpha_s^3$, the exact result is recovered.
Thus the use of $Y_{\rm eff}$ instead of $Y$ in the BFKL expression can correct in an effective way the potentially too strong assumptions made in a BFKL calculation while preserving the additional emissions of gluons specific to this approach.
The value of $Y_{\rm eff}$ is an indication of how valid the BFKL approximation is: a value close to $Y$ means that this approximation is valid, whereas a value significantly different from $Y$ means that it is a too strong assumption in the kinematics under study.

\psfrag{LO}[l][l][.8]{LL}
\psfrag{NLO}[l][l][.8]{NLL}
\psfrag{yeffy}[l][l][.8]{$Y_{\rm eff}/Y$}
\psfrag{kj1}[l][l][0.8]{\hspace{0.1cm}$\veckjtwo$ (GeV)}
\begin{figure}[h]
\centering\includegraphics[width=7.5cm]{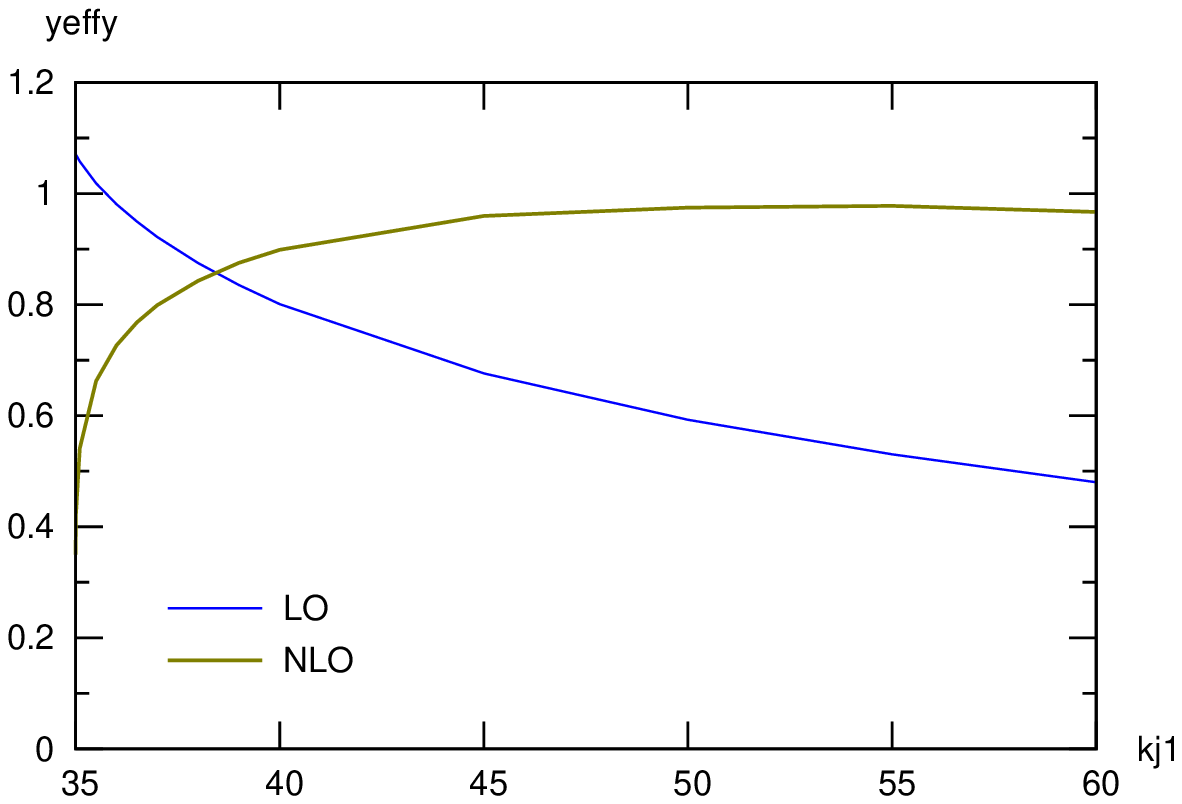}
\caption{Variation of $Y_{\rm eff}/Y$ as defined in eq.~(\protect\ref{eq:Cm_e-m_cons})
as a function of $\veckjtwo$ at fixed $\veckjone=35$ GeV for $Y=8$ and $\sqrt{s}=7$ TeV at leading logarithmic (blue) and next-to-leading logarithmic (brown) accuracy.}
\label{Fig:yeff_NLO}
\end{figure}

On fig.~\ref{Fig:yeff_NLO} we show the values obtained for $Y_{\rm eff}$ as a function of $\veckjtwo$ for fixed $\veckjone=35$ GeV at a center of mass energy $\sqrt{s}=7$ TeV and for a rapidity separation $Y=8$, in the LL and NLL approximation. We see that, as found in ref.~\cite{DelDuca:1994ng}, the LL calculation strongly overestimates the cross section when the transverse momenta of the jets are not very similar.
This is no longer the case at NLL accuracy: now, when the transverse momenta of the jets are significantly different (as needed to obtain trustable results in the fixed order approach), the effective rapidity is very close to $Y$ meaning that the violation of energy-momentum should be much less severe at NLL accuracy.

\section{Conclusions}

We have studied the azimuthal correlations of Mueller-Navelet jets
and compared the predictions of a full NLL BFKL calculation with data taken at
the LHC. We have shown that the use of the BLM procedure to fix the renormalization
scale leads to a very good agreement with the data, which is much more satisfactory than when using
the 'natural' value $\sqrt{|\veckjone|\cdot|\veckjtwo|}$.
We also studied the effect of the absence of strict energy-momentum conservation in a BFKL calculation, and showed that for significantly different values of transverse momenta of the tagged jets this effect is expected to be tiny at NLL accuracy.

\acknowledgments

We thank  Stan Brodsky, Grzegorz Brona, Michel Fontannaz, Tomasz Fruboes, Hannes Jung, Victor Kim, Cyrille Marquet and Maciej Misiura for fruitful discussions.

This work is supported by the French Grant  PEPS-PTI,
the Polish Grant NCN No.~DEC-2011/01/B/ST2/03915 and  the Joint Research Activity 
Study of Strongly Interacting Matter (HadronPhysics3, Grant Agreement n.283286)
under the 7th Framework Programme of the European Community.

\providecommand{\href}[2]{#2}\begingroup\raggedright\endgroup

\end{document}